\documentclass[aps,amssymb,amsmath, twocolumn, pra, superscriptaddress]{revtex4}
\usepackage{graphicx}
\usepackage{epsfig}
\usepackage{dcolumn}
\usepackage[up]{subfigure}
\usepackage{float}
\usepackage{dsfont}	
\usepackage{relsize}

\DeclareMathOperator{\Tr}{Tr}
\usepackage{epstopdf}

\usepackage{float}
\usepackage{ragged2e}
\usepackage{braket}
\usepackage[font=small,labelfont=bf,justification=raggedright, format=plain]{caption}
\usepackage{setspace}
\usepackage[colorlinks, citecolor = magenta]{hyperref}

\begin{document}

%======================================================================================

\title{Noise resilience in path-polarization hyperentangled probe states}

\author{Akshay Kannan Sairam}
\email{akshaykannan@iisc.ac.in}
\affiliation{Quantum Optics \& Quantum Information, Department of Instrumentation and Applied Physics, Indian Institute of Science, Bengaluru, India}
\author{C. M. Chandrashekar}
\email{chandracm@iisc.ac.in}
\affiliation{Quantum Optics \& Quantum Information, Department of Instrumentation and Applied Physics, Indian Institute of Science, Bengaluru, India}
\affiliation{The Institute of Mathematical Sciences, C. I. T. Campus, Taramani, Chennai 600113, India}
\affiliation{Homi Bhabha National Institute, Training School Complex, Anushakti Nagar, Mumbai 400094, India}

%======================================================================================
\begin{abstract}
Most quantum systems that are used for generating entanglement and for practical applications are not isolated from the environment, and are hence susceptible to noise. Entanglement in more than one degree of freedom between two systems, known as hyperentanglement, is known to have certain advantages, including robustness against noise over conventional entangled states.   Quantum illumination, imaging and communication schemes that involve sending one photon from a pair of entangled photons and retaining the other photon usually involve exposing only the signal photon to environmental noise. The disruptive nature of noise degrades entanglement and other correlations which are crucial for many of these applications. In this paper, we study the advantages of using photon pairs in certain path-polarization hyperentangled states in a noisy interaction where photons in only one of the paths are affected by noise. We model such noise and study the effect of noise on the correlations present in the hyperentangled photons. Three different methods, entanglement negativity, entanglement witnesses and Bell nonlocality are used to show the resilience of path-polarization hyperentangled probe state against noise.
\end{abstract}

%======================================================================================

\maketitle

%======================================================================================
\section{Introduction}

%======================================================================================
Quantum entanglement is a useful resource in many domains of science which harness the exotic effects of quantum mechanical systems for better performance or security in information processing, metrology and communication tasks\,\cite{Entanglement_Horodecki,Advances2011Giovanetti}. Some examples of the pioneering applications of entanglement that beat the best known classical protocols are quantum dense coding, quantum teleportation, quantum key distribution, and quantum illumination \cite{Bennet1992Dense,Teleporting1993Bennett,E91,LLoyd_QI2008}. Entanglement is also deemed as one of the key resources behind computational speedup in quantum computing \cite{Josza2003Role}. It is an essential ingredient in device independent quantum cryptography \cite{Acin2007Deviceindependent}. Entanglement in higher dimensions has been shown to give better performance enhancements than conventional entanglement, like enhanced channel capacity and robustness against noise in quantum communication \cite{Hu2020Beating,Collins2002Arbitrary,Cerf2002Security,Zhu2020IsHigh}.

Entangled photons are widely used in many applications since they can be readily generated through non-linear processes like Spontaneous Parametric Down Conversion (SPDC) using birefringent crystals, and transmitted relatively easily \cite{NewSourceEntanglement_Kwiat1995}. Interactions with the environment can subject the quantum system to noise, which leads to a loss of entanglement, decoherence and other such disruptive effects. Photonic dissipation in waveguide quantum electro dynamical (QED) systems is another example of effect of environmental interaction \cite{Chen2017Exact,Chen2017Dissipation,Chen2018Entanglement}. In some applications, the transmission of one of the entangled photon pair and retention of the other, causes environmental noise to act only on the transmitted photons in the noisy path. Such a scheme is commonly adopted for example in quantum cryptographic protocols, quantum illumination schemes, and quantum teleportation protocols \cite{E91,LLoyd_QI2008}. Therefore, finding ways to engineer entangled states that are robust against noise is an area of continuous research interest. 

\textit{Hyperentanglement---} Simultaneous entanglement between two systems in more than one degree of freedom, is termed as hyperentanglement \cite{Kwiat1996Hyperentanglement}. Photons with more than one controllable degree of freedom can be simultaneously entangled in multiple degrees of freedom like path, polarization, time, energy and orbital angular momentum \cite{GenerationHyper_Barreiro2005,PMHyper_Barbieri2005,BCM05, BMM06, Zhao2019Direct, Dong2014Generation, Yaasir2021Generation}.  The presence of entanglement in more than one degree, expands the dimension of the Hilbert space of the state of the photon pair. Some of the applications that utilise hyperentangled states are enhanced channel capacity, protocols for entanglement purification, quantum secure direct communication and enhanced signal to noise ratio in quantum illumination.\cite{Barreiro2008Beating,Sheng2021Onestep, Hu2021LongDistance, Huang2022Experimental, HyperentanglementQI_AVPrabhu2021}.  In this work we will consider one such configuration of hyperentangled state,  entanglement in path and polarization degree of freedom and show its advantages in the desired noisy environment where noise acts only on one of the paths. We use controlled Kraus operators of bit flip, phase flip and depolarizing channel to model the noise. Different indicators of correlations, which are entanglement negativity, entanglement witnesses, and Bell nonlocality after the action of noise are used to quantify and compare the results with conventional entangled photons. It can be seen that the path-polarization hyperentangled state retains correlations as much or better than conventional entangled photons when affected by noise.

The paper is organised as follows. In Sec.\,\ref{sec:noise_model}  a noise model is introduced that can be used to study scenarios like quantum cryptography and quantum illumination. In Sec.\,\ref{sec:Quant_Correlations} we introduce the quantifiers of correlations we use to study the effect of noise. In Sec.\,\ref{sec:results} we present the numerical results showing the robustness of path-polarization hyperentangled state against noise when compared to entanglement in one degree of freedom and conclude with remarks in Sec.\,\ref{conc}.

%======================================================================================
\section{Path-polarization entanglement and noise model}
\label{sec:noise_model}
%======================================================================================
Photons in composition of polarization and path degree of freedom will have a Hilbert space composition, $\mathcal{H}_{pol} \otimes \mathcal{H}_{p}$ where $\mathcal{H}_{pol}$ is spanned by the basis state $\{|H\rangle, |V\rangle \}$ representing the horizontal and vertical polarization degree of freedom, and  $\mathcal{H}_{p}$ is spanned by the basis states $\{|0\rangle, |1\rangle \}$ representing the two paths for photons. The two paths can be, for example, the output modes of a beam splitter. Path-polarization hyperentangled state can be generated by first generating photons entangled in polarization degree of freedom using SPDC process and by further engineering photons paths to entangle in path degree of freedom \cite{BCM05, BMM06}. Appropriate post selection of the states can also be used to generate path-polarization hyperentangled states \cite{PMHyper_Barbieri2005}. In this work, we will model the effect of noise on such hyperentangled states where the noise acts only on the photons present in one of the path,  $\ket{0} \in \mathcal{H}_p $ and leaves the photon in the other path unaffected.  Such noise can be mathematically described in the form of a controlled noise model  in resemblance to the controlled unitary operator where noise acts on the target (polarization) system conditioned on the state of the control (path) system. 

In order to model noise in such a way, it is required that a few conditions are met. For a photon state that is coupled with the spatial mode $\ket{0}_1\ket{0}_2$, the noise should act on both the photons. Similarly, for photon states having spatial modes, $\ket{0}_1\ket{1}_2, \ket{1}_1\ket{0}_2$, only single photon noise should act on the photon in $\ket{0}$, and the corresponding polarization entangled photon should remain unaffected by the noise. Finally, for photons passing through $\ket{1}_1\ket{1}_2$, both the photons should remain unaffected. Such situations can often be seen in applications like quantum illumination \cite{LLoyd_QI2008}, where one of the noisy paths is probed for an object while the other noiseless path is kept as a reference. Similar situations also arise in quantum teleportation and quantum key distribution, where one of the photons of an entangled photon pair is sent to Bob and hence subject to noise, while storing the other photon as a reference \cite{Teleporting1993Bennett,E91}. 

We model noise using Kraus operator representation of noise channels. These are a set of operators that are derived from jointly evolving a state along with an environment and tracing out the environment \cite{nielsenchuang}. Given a set of Kraus operators describing a single photon noise model, like bit flip, depolarizing and phase damping noise, 

\begin{equation}
 \label{eq:kraus} 
K^1_i \equiv \{K_1, K_2,\cdots K_n\}.
\end{equation} 
The Kraus operators for two photons can be constructed by taking tensor products of combinations of the set of single qubit Kraus operators as shown in Eq.(\ref{eq:twoqubit}). This parallel concatenation of multiple channels can be generalized for higher dimensional composite photon states \cite{mmwilde_qiqc}. Here we construct the Kraus operators for the two photon channel corresponding to Eq.(\ref{eq:kraus}).
\begin{equation}
\label{eq:twoqubit}  
K^2_i \equiv \{ K_i\otimes K_j :K_i,K_j \in \{K^1_i\} \}.
\end{equation}

To construct the noise model, the single/two photon noise Kraus operators are coupled with the appropriate projectors from a set of projection operators on the position Hilbert space, 
\begin{align}
 \label{eq:projectors} 
 P_{00} &= \ket{0}_1\ket{0}_2\bra{0}_1\bra{0}_2\\\nonumber
 P_{01} &= \ket{0}_1\ket{1}_2\bra{0}_1\bra{1}_2\\ \nonumber
 P_{10} &= \ket{1}_1\ket{0}_2\bra{1}_1\bra{0}_2\\ \nonumber
 P_{11} &= \ket{1}_1\ket{1}_2\bra{1}_1\bra{1}_2,   \nonumber
\end{align}
and put together as a single set of combined Kraus operators $\tilde{K_i}$,
\begin{equation}
\label{eq:controlled_kraus}  
\tilde{K_i} \equiv
\left\lbrace
\begin{aligned}
              & \begin{aligned}&(K^1_i\otimes I) \otimes P_{01},\\&(I\otimes K^1_i) \otimes P_{10},\end{aligned} &\text{(Single photon noise)} \\
              & (K^2_i)\otimes P_{00}, &\text{(Two photon noise)} \\
              &(I\otimes I) \otimes P_{11} &\text{(Identity channel)}
\end{aligned}
\right\rbrace
.
\end{equation}
It can be verified that the action of the above Kraus operators satisfies all the required conditions of a quantum channel, including the Complete Positivity and Trace Preserving (CPTP) condition. We note that this noise model does not accurately model the noise acting on the path states, since the projectors cause loss of correlations in path degree of freedom. But in this study we are examining only the polarization degree of freedom, hence this model is valid.

The noise model developed above is applied on the hyperentangled state of the form,
\begin{align}
 \label{eq:Hyperentangled}
 \ket{\Psi}_{HE} = \frac{1}{2}(\ket{H}_1\ket{V}_2 + \ket{V}_1\ket{H}_2) \otimes (\ket{1}_1\ket{1}_2 + \ket{0}_1\ket{0}_2).
\end{align}
Here subscript 1 and 2 represent photon 1 and photon 2. In the above expression we can note that both the photons entangled in polarization degree of freedom take the same path in this configuration. However, the polarization degree entangled photons will remain spatially separated along each path. Such states can be generated in a laboratory setting by using the methods illustrated in references \cite{PMHyper_Barbieri2005,Simon2002Polarization}. Conventional polarization entangled photons that are coupled with only one of the spatial modes is used as a reference. For ease of comparing these states, the spatial mode of the conventional entangled photon pair is taken to be $\ket{0}_1\ket{1}_2$. Now the path-polarization entangled and polarization entangled states belonging to $\mathcal{H}_{pol} \otimes \mathcal{H}_{p}$ can be written as,

\begin{align}
  \label{eq:path_and_reference}
\ket{\Psi}_{HE} =
 \frac{1}{2}&(\ket{HV00} + \ket{HV11} + \ket{VH00}) + \ket{VH11}),\\
 \ket{\Psi}_E = \frac{1}{\sqrt{2}}&\left(\ket{HV01} + \ket{VH01}\right).
\end{align}
The Kraus operators act on the density matrices of the photon states. The corresponding density matrix representations of the photon states are,
\begin{equation}
\label{eq:density_matrix} 
\rho_{HE} = \ket{\Psi_{HE}}\bra{\Psi_{HE}}, \quad \rho_E = \ket{\Psi_E}\bra{\Psi_{E}}. 
\end{equation}
The action of the noise model is then given by,
\begin{equation}
\label{eq:noise_action}
\tilde{\rho}_{\mbox{\tiny(HE)}} = \sum_i \tilde{K_i} \rho_{(\mbox{\tiny HE)}} \tilde{K_i}^{\dagger}, \quad 
\tilde{\rho}_{\mbox{\tiny(E)}} = \sum_i \tilde{K_i} \rho_{(\mbox{\tiny E)}} \tilde{K_i}^{\dagger}.
\end{equation}

Following this, the path degree of freedom is partial traced out for both the states bringing down both of these states to a 4-dimensional space. This is done in order to compare between the entanglement in both these pairs of photons on an equal footing.

\begin{equation}
\label{eq:part_trace}
\rho^{\text{out}}_{\mbox{\tiny HE}} = \text{Tr}_{\mbox{\tiny POS}}(\tilde{\rho}_{\mbox{\tiny(HE)}}),\quad
\rho^{\text{out}}_{\mbox{\tiny E}} = \text{Tr}_{\mbox{\tiny POS}}(\tilde{\rho}_{\mbox{\tiny(E)}}). 
\end{equation}

The controlled noise model can be used with any set of single qubit Kraus operators by substituting in place of Eq.(\ref{eq:kraus}) and getting the corresponding set of controlled Kraus operators $\tilde{K}$.

%=======================================================
\section{Quantifying Correlations under noise}
\label{sec:Quant_Correlations} 
%=======================================================

There are many measures that can reliably quantify (quantum) correlations in a composite system \cite{Olivier2001Discord,Hill1997Entanglement,ComputableEntanglment_Vidal2002}.  To study the effect of noise on the quantum correlations of the photon states, we introduce an entanglement measure called negativity and a measure of nonlocality using the CHSH parameter. We will also take a look at entanglement witnesses which present an effective method to experimentally verify the presence of entanglement in the system. 

%======================================================================================
\subsection{Entanglement Negativity}
\label{subsec:EntanglementN}
%======================================================================================

Given a density matrix of a composite quantum system, the entanglement negativity $\mathcal{N}$ between two bipartitions A:B of the system is given as,
\begin{equation}
  \label{eq:negativity}
  \mathcal{N}(\rho) = \frac{||\rho^{\Gamma_A}||_1 - 1 }{2},
\end{equation}
where $||\rho^{\Gamma_A}||$ denotes the partial transpose of $\rho$ with respect to subsystem A, and $||X||_1 = \text{Tr}(\sqrt{XX^{\dagger}})$ denotes the trace-norm \cite{ComputableEntanglment_Vidal2002,VolumeofSeprablestates_Zyckowski1999}. For the path encoded state, the polarization degree of freedom of the two photons are considered as the subsystems. For the purpose of comparing conventional entangled photons and path-encoded photons, we trace out the position degree of freedom of the path-encoded state after applying the operations to it. 
Initially, it can be seen that the hyperentangled states have entanglement negativity of $\mathcal{N} = 1$. The state of the hyperentangled photons in position degree of freedom contributes equally to the entanglement negativity as the polarization and can be added up using the additive property of negativity. 

\textit{Entanglement Witness---}
Experimentally, it is difficult and resource intensive to measure entanglement negativity without knowing the complete density matrix of the state through methods like quantum state tomography (QST) \cite{QST_DanielFV2001}. For this reason, we construct an entanglement witness, which are operators whose expectation value can be used as an indicator for whether the state is entangled or not \cite{EntanglementWitness_Terhal2000}. A state $\rho$ is entangled if and only if there exists a Hermitian operator W such that $\text{Tr}(\rho W)<0$ and $\text{Tr}(\rho_{sep} W)\geq 0$ for all separable states $\rho_{sep}$. The operator W can then be defined as the entanglement witness of the state $\rho$. The expectation values of W can be measured experimentally using fewer number of measurements as compared to QST. We study here theoretically how the expectation values of suitable witness operators are affected by various levels of noise, and illustrate a method to experimentally measure $\langle W \rangle$ using fewer measurements \cite{DetectionofEntanglment_Guhne2002}.  It should be noted that given an entanglement witness for a particular entangled state, it may not be a suitable witness for other entangled states.

%=================================================
\subsection{Nonlocality Measurments}
\label{sec:nonlocal}
%===============================================
A useful benchmark for certification of entanglement in a system is the violation of the Bell's inequality \cite{Bell1964}. (Clauser Horne Shimony Holt) CHSH inequality measurements can be performed in laboratory settings using an entangled photons pairs and local measurements using optical components like half-wave plates or polarizers, and coincidence detections \cite{CHSH_aspect1981,CHSH_1970}. We investigate the effect of the higher dimensional entanglement on the CHSH parameter S. 

Given a bipartite quantum system, we can define two parties, Alice and Bob, who perform local measurements on one of the photons. They can choose from a set of orthogonal measurements independently. The CHSH inequality is given by,
\begin{equation}
 \label{eq:CHSH} 
S = E(a,b) - E(a,b') + E(a',b) + E(a',b').
\end{equation}
Here $a, a'$ and $b, b'$ represent the measurements with outcomes 1 or -1, taken by Alice and Bob respectively. E(a,b) represents the expectation value for the measurement a and b. According to Bell's theorem of nonlocality, for physical systems that can be described by local hidden variable theories, the CHSH inequality is given by,
\begin{equation}
\label{eq:bound} 
|S|\leq 2.
\end{equation}

Quantum systems are known to violate the CHSH inequality, indicating the nonlocality of quantum mechanics. Here we theoretically investigate the Bell nonlocality of hyperentangled photons under the effect of noise. As a reference, we also see the effect of noise on the nonlocality of conventional photons. In the context of photons, $a, a'$ and $b, b'$ are determined by the angles $\theta, \theta'$ and $\delta, \delta'$ for along which the polarization ($\ket{H}$ or $\ket{V}$) of the photon are measured. 

In the experimental setting, measurements are made in the form of coincidence counts using suitable time tagging devices. The measurements are performed for four different combinations of polarization of photons: $C_{HH}, C_{HV}, C_{VH}, C_{VV}$. Here the $C_{HV}$ stands for coincidence counts when first photon is measured for the polarization $H$ and the second photon measured for polarization $V$. Using these coincidence counts, $E(\theta,\delta)$ can be obtained using, 
\begin{equation}
  \label{eq:E_experiment}
E(\theta,\delta) = \frac{C_{HH} + C_{VV} - C_{HV} - C_{VH}}{C_{HH} + C_{VV} + C_{HV} + C_{VH}},
\end{equation}
where $\theta$ and $\delta$ denote the angle of rotation of the polarization of two photons at which the coincidence counts are measured for each of the photon pair respectively. S can now be computed using the value of $E$ for different combinations of rotation angles as given in Eq.(\ref{eq:CHSH}). 

Theoretically, the S parameter can be calculated by obtaining the probability of measurements for various configurations, by either performing rotation operations on the density matrix or performing a measurement using a rotated projection operator. For the first case,
\begin{align}
  \label{eq:CHSH_measure_theoretical} 
  \rho^{out}_{rot} &= (R(\theta) \otimes R(\delta))\;\rho^{out}_{(E/HE)}\; (R(\theta)^{\dagger}\otimes R(\delta)^{\dagger}),\\ \nonumber
  P_{VV} &= \Tr(\mathcal{P}_{VV}\rho^{out}_{rot}).
\end{align}
Where R($\theta$) is the rotation operator acting on the polarization degree of freedom given by, 

\begin{equation}
	R(\theta) = 
	\begin{pmatrix}
	\cos(\theta)&-\sin(\theta)\\
	\sin(\theta)&\cos(\theta)	
	\end{pmatrix},
\end{equation}

and $ \mathcal{P}_{VV}$ denotes the projection operator along the state $\ket{VV}$. Now,
\begin{equation}
  \label{eq:E_theoretic} 
E(\theta,\delta) = {P_{HH} + P_{VV} - P_{HV} - P_{VH}}.
\end{equation}

\begin{figure}[h!]
  \centering
\includegraphics[scale = 0.5]{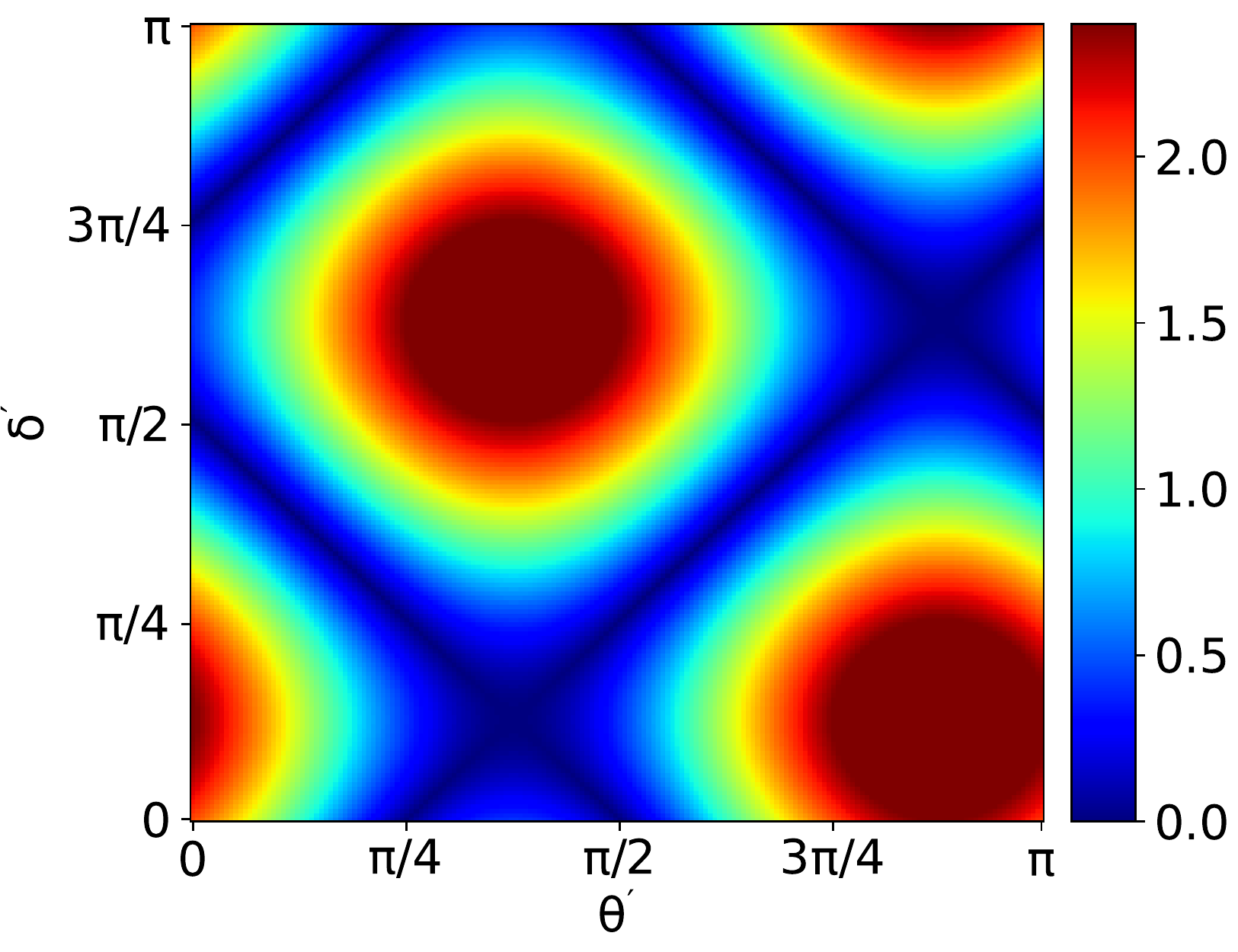}
\caption{CHSH parameter $S$  for the polarization entangled state as a function of $(\theta^{\prime}, \delta^{\prime})$ when  $(\theta, \delta)  = (\pi/4,  \pi/2)$. The maximum violation of CHSH inequality, $S=2\sqrt 2$ is seen when $(\theta^{\prime}, \delta^{\prime}) = (3\pi/8, 5\pi/8)$ and $(\theta^{\prime}, \delta^{\prime}) = (7\pi/8, \pi/8)$.
 \label{fig:refnon}} 
\end{figure}

In Fig.\,\ref{fig:refnon}, in absence of noise, the CHSH parameter $S$ is calculated and plotted for all angles of $\theta'$ and $\delta'$ in the range $0\leq\theta'\leq\pi$, $0\leq\delta'\leq\pi$  when $\theta = \pi/4$ and $\delta = \pi/2$.  For $(\theta^{\prime}, \delta^{\prime}) = (3\pi/8, 5\pi/8)$ and $(\theta^{\prime}, \delta^{\prime}) = (7\pi/8, \pi/8)$  when $(\theta, \delta)  =(\pi/4, \pi/2)$ we obtain a maximum violation of CHSH inequality, $S = 2\sqrt 2$. We will use this as a reference to see the effect of noise on the maximum value of $S$.

%==========================
\section{Results}
\label{sec:results}
%==========================

In this section, we study the effect of the noise modeled using the Eq.(\ref{eq:controlled_kraus}) on the correlations of the photon states. Three different basic noise models are used here \cite{nielsenchuang}: (1) bit flip (2) depolarizing (3) phase damping. We numerically plot the variation of the quantifiers introduced in Sec.\,\ref{sec:Quant_Correlations} as a function of the noise levels and compare the effect of noise on hyperentangled states and entangled states.

\subsection{Bit Flip Noise}
The bit flip noise channel, as the name suggests, is used to model bit flip errors in two level quantum systems. The noise parameter p in the context of bit flip noise is the probability of a flip ($\ket{H} \leftrightarrow \ket{V}$) occurring. The Kraus operators for the single qubit flip noise channel are given by,

\begin{equation}
K_1 = (\sqrt{p})\begin{pmatrix}
0 & 1 \\
1 & 0 
\end{pmatrix},
\quad
K_2 = (\sqrt{1-p})\begin{pmatrix}
1 & 0 \\
0 & 1 
\end{pmatrix}.
\end{equation}

On the action of the bit flip channel on the hyperentangled photons and the entangled reference states, the output state $\rho^{\text{out}}_{\mbox{\tiny(E/HE)}}$ is computed for various values of p. Here, p quantifies the amount of noise being acted on the photon states. $\mathcal{N}(\rho^{\text{out}}_{\mbox{\tiny(HE)}})$ and $\mathcal{N}(\rho^{\text{out}}_{\mbox{\tiny(E)}})$ can now be computed using Eq.(\ref{eq:negativity}).

\begin{figure}[h!]
\includegraphics{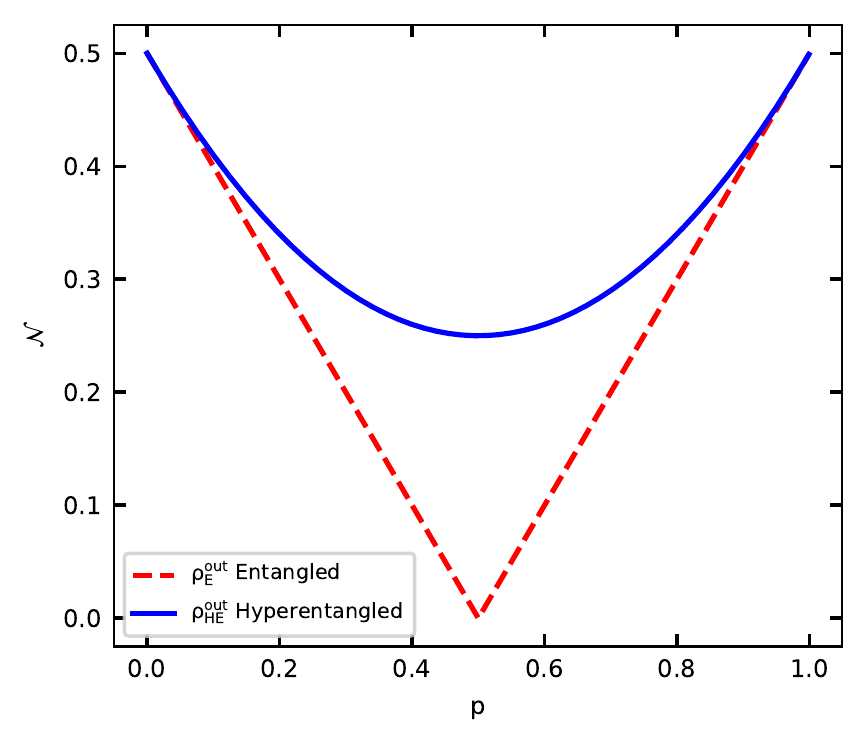}
\caption{Entanglement negativity, $\mathcal{N}$ as a function of bit flip noise level $p$. The plot for the hyperentangled state shows a consistently higher level of negativity than the plot for the entangled state
indicating the robustness of the hyperentangled state against bit flip noise.}
\label{fig:blipcontrol}
\end{figure}

In Fig.\,\ref{fig:blipcontrol}, the plot for negativity as a function of noise level $p$ shows enhanced retention of entanglement in hyperentangled states. The enhancement can be seen as a result of the state that was chosen that includes a superposition of the noisy and noiseless path, which reduces the effect of noise acting on the complete state. Since the bit flip noise acts symmetrically on the polarization state of the two photons, the maximum noise level will be $p=0.5$.  It can be observed that the value of negativity returns back to the highest value of $\mathcal{N} = 0.5$ with the increase in $p$ from $0.5$ to $1$ indicating the return to the maximally entangled two photon (Bell) state.

\begin{figure}[h!]
 \includegraphics{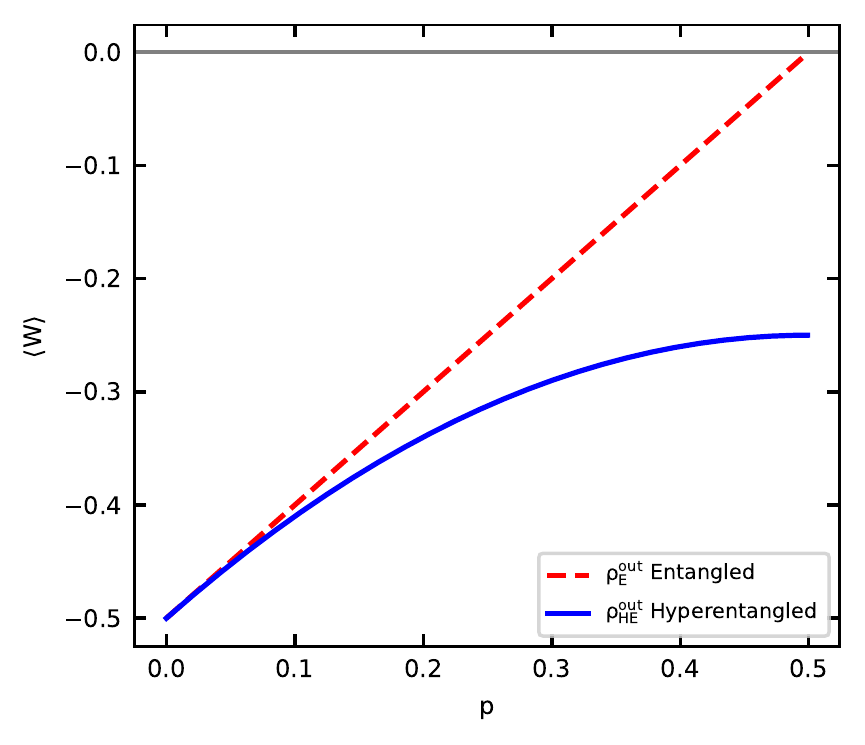}
 \caption{Entanglement witness $\langle W \rangle$ as a function of bit flip noise in the range,  $0\leq p < 0.5$. The expectation value of the witness for hyperentangled state has a negative value at maximum bit flip noise level, $p =0.5$ where the the expectation value of witness for the entangled state becomes $\langle W \rangle = 0$ indicating the state becoming separable.}
 \label{fig:witnessblip}
\end{figure}

Now for an entanglement witness as explained in the Sec.\,\ref{sec:Quant_Correlations}, for the bit flip channel (with $0 \leq p \leq 0.5 $ ) there exists a simple witness,
\begin{equation}
 \label{eq:witness} 
 W = \frac{1}{2}
\begin{pmatrix}
   1 & 0 & 0 & 0\\ 
   0 & 0 & -1 & 0\\ 
   0 & -1 & 0 & 0\\ 
   0 & 0 & 0 & 1
\end{pmatrix}.
\end{equation} 
The operator $W$ can be easily decomposed to a set of local measurements \cite{DetectionofEntanglment_Guhne2002},
\begin{equation}
 \label{eq:witness_decomp} 
 W = \frac{1}{2}(\sigma_i \otimes \sigma_i  - \sigma_x \otimes \sigma_x - \sigma_y \otimes \sigma_y + \sigma_z \otimes \sigma_z).
\end{equation}
Such a measurement can be performed in the laboratory using local measurements in the Pauli basis, and combining the results with the weights.  In the case of QST, expectation values of 16 operators \{$\sigma_i\otimes\sigma_j:i,j = 0,1,2,3\}$ are required, but in the case of the entanglement witness W, only 4 of them are required. The theoretical expectation value of $W$ is plotted in Fig.\,\ref{fig:witnessblip}. In the following subsections, we see that the same entanglement witness, Eq.($\ref{eq:witness}$) can be used as a witness for other noise models discussed in this work.

In the Fig.\,\ref{fig:witnessblip}, it can be seen that both plots approach the value $\langle W \rangle$ with increasing levels of noise and at p = 0.5, the witness for the entangled state reaches $\langle W \rangle = 0$ indicating separability ($\mathcal{N} = 0$), but the hyperentangled state has negative value of  $\langle W \rangle$ indicating entanglement, which is in agreement with the negativity plot Fig.\,\ref{fig:blipcontrol}. Although $\langle W \rangle$ does not explicitly quantify entanglement, it can be used to comparatively illustrate the robustness of the states against noise.

\begin{figure}[h!]
\includegraphics[scale = 0.32]{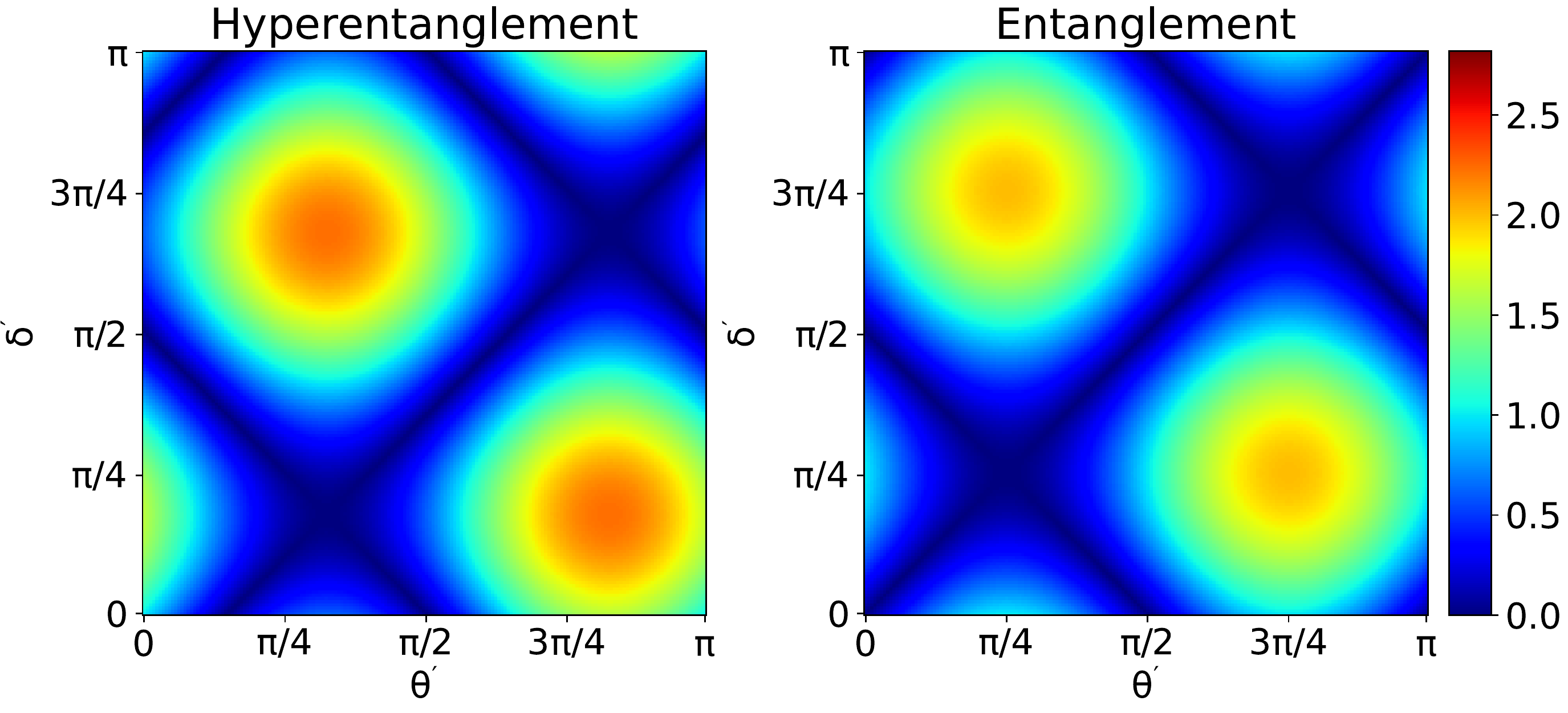}
\caption{CHSH parameter $S$ as a function of $\theta^{\prime}$ and $\delta^{\prime}$ when $(\theta, \delta)  =(\pi/4, \pi/2)$ and bit flip noise level $p=0.5$.  It is evident that the bit flip noise has degraded the nonlocality effect in both states when compared with Fig.\,\ref{fig:refnon} but the maximum value of $S$ is seen to be higher for the hyperentangled state.
 \label{fig:blipnon}} 
\end{figure}

\begin{figure}[h!]
 \centering
\includegraphics{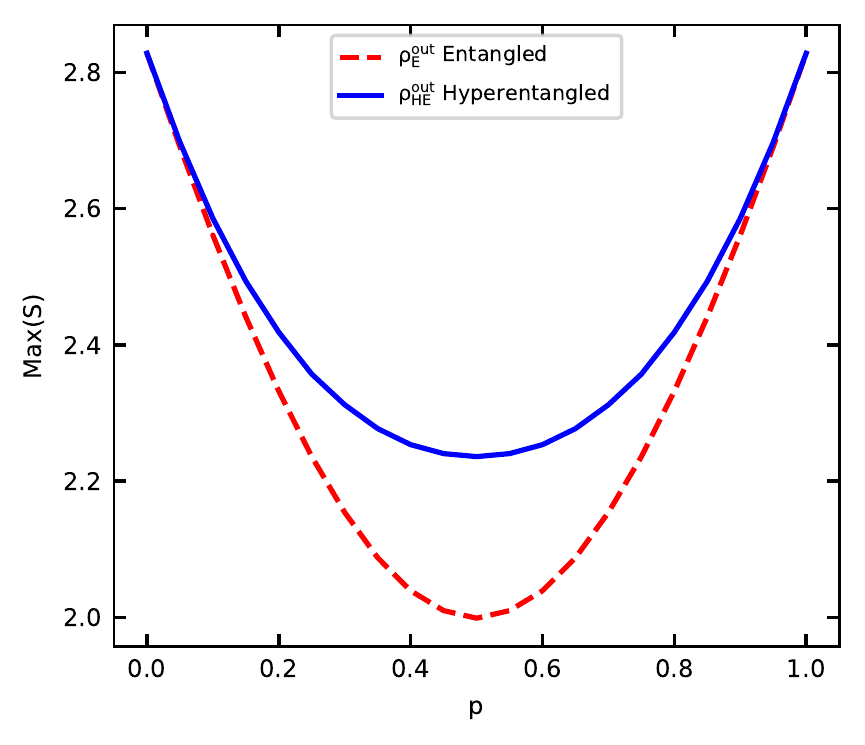}
\caption{Maximum value of CHSH parameter $S$ as a function of bit flip noise level $p$ when $(\theta, \delta) = (\pi/2, \pi/4)$ and $\theta',\delta' \in [0,\pi]$. Max(S) for the given range of angle configurations seem to follow a similar trend to that of negativity. It is evident that hyperentangled photons are able to retain nonlocality better than entangled photons.}
 \label{fig:blipnonplot} 
\end{figure}

For the nonlocality measurements we compute the theoretical value of the CHSH parameter $S$ as explained in Sec.\,\ref{sec:nonlocal}. The S($\theta',\delta'$) value for $\theta = \pi/4$ and $\delta = \pi/2$ is computed for $\rho^{out}_{E}$ and $\rho^{out}_{HE}$ for the bit flip channel with a fixed value of p = 0.5. There is a visible difference in the pattern between $\rho^{out}_{E}$ and $\rho^{out}_{HE}$ in the Fig.\,\ref{fig:blipnon}, but it does not give much insight into the difference between the two channels. However, there is no violation of CHSH inequality in this range indicating that the noise has degraded the nonlocality of both the states. 
It is quantitatively clear from the plots that when noise level, $p=0.5$ the effect of noise on nonlocality, CHSH parameter $S$ is higher for the entangled state compared to the hyperentangled state. For a better comparison of the nonlocality of $\rho_{HE}$ and $\rho_{E}$ the maximum value of S in the range $0\leq\theta',\delta'\leq\pi$ with $\theta = \pi/2$ and $\delta = \pi/4$ is plotted in Fig.\,\ref{fig:blipnonplot} as a function of noise level $p$. Similar to negativity, the nonlocality appears higher for the hyperentangled photons. In an experimental setting it may be difficult to iterate over all the possible angles to compute the S value and find the maximum value, but the plots in  Fig.\,\ref{fig:refnon} and other such plots will be a good reference to identify the regions where maximum $S$ can be obtained.

%==================================
\subsection{Depolarizing Noise}
\begin{figure}[h!]
\includegraphics{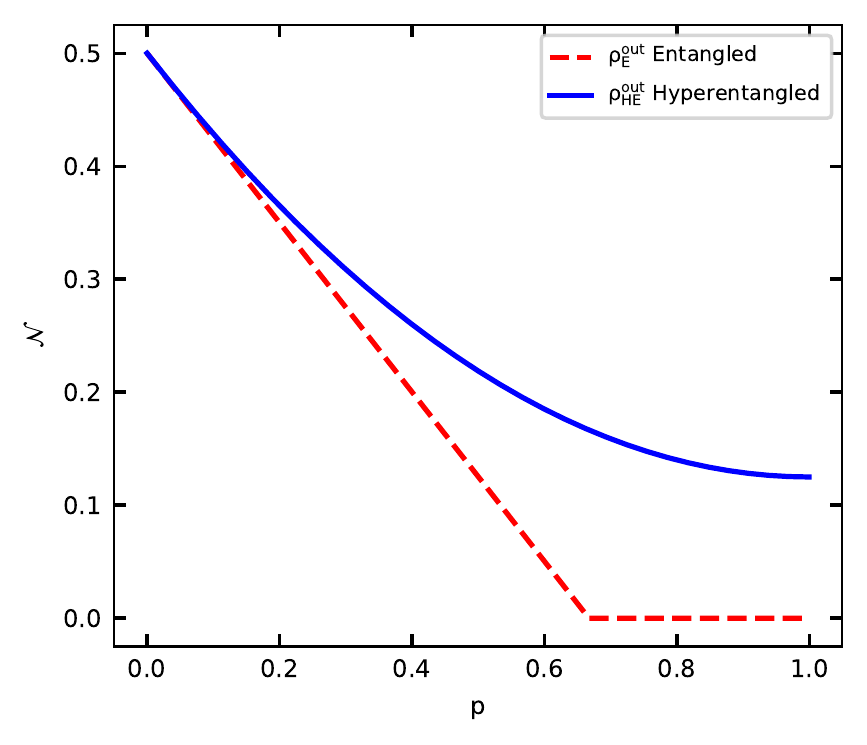}
\caption{Entanglement negativity $\mathcal{N}$ as a function of depolarizing noise level $p$. The negativity of the entangled state reaches zero faster than the hyperentangled state, indicating the robustness of the hyperentangled state against depolarizing noise.}
\label{fig:depolcontrol} 
\end{figure}

\begin{figure}[h!]
  \includegraphics[scale = 1]{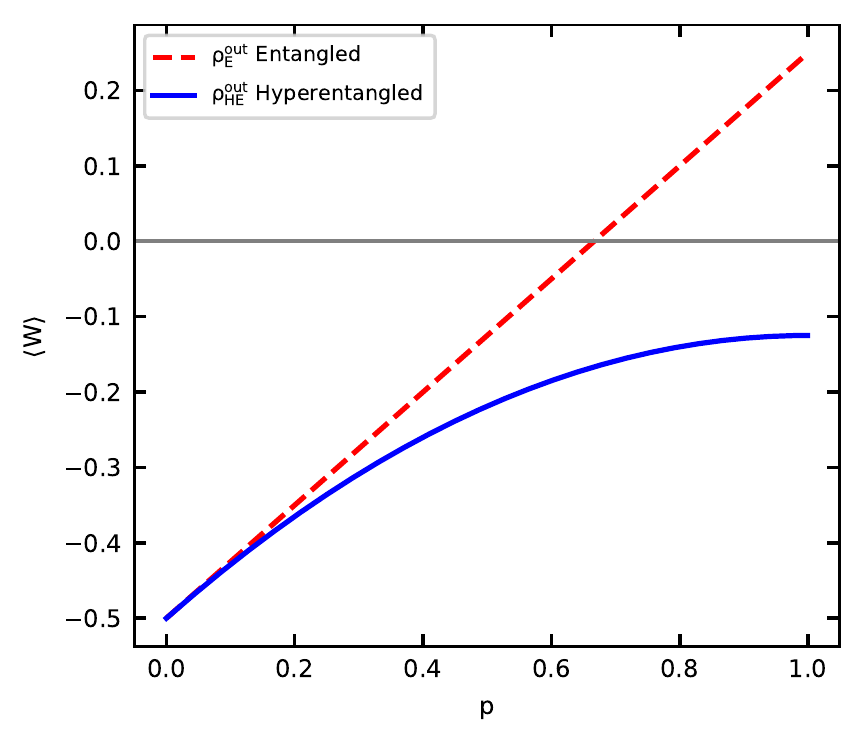}
  \caption{Entanglement witness $\langle W \rangle$ as a function of depolarizing noise level $p$. The plot for the entangled state reaches $\langle W \rangle = 0$ at the same level of $p$ for which $\mathcal{N} = 0$, and the plot for the hyperentangled state remains in the negative region for all levels of noise.}
  \label{fig:witnessdepol}
\end{figure}

%===================================
Depolarizing noise channel is a combination of bit flip, phase flip, and bit plus phase flip operators (represented by Pauli matrices). The noise parameter here serves as the probability of the occurrence of flips. The single qubit Kraus operators for the depolarizing noise channel are, 
\begin{equation}
 \label{eq:depolcontrol} 
\begin{aligned}
&K_1 = \sqrt{p/4}\begin{pmatrix}
0 & 1 \\
\hspace{2mm}1 & 0 
\end{pmatrix}
\quad
&K_2 = \sqrt{p/4}\begin{pmatrix}
1 & 0 \\
0 & -1 
\end{pmatrix}\\[0.5cm]
&K_2 = \sqrt{p/4}\begin{pmatrix}
0 & i \\
-i & 0 
\end{pmatrix}
\quad
&K_4 = \sqrt{1-3p/4}\begin{pmatrix}
1 & 0 \\
0 & \hspace{2mm}1 
\end{pmatrix}.
\end{aligned}
\end{equation}
In Fig.\,\ref{fig:depolcontrol} and Fig.\,\ref{fig:witnessdepol} the negativity and witness as a function of depolarizing noise level is shown. One can clearly see the robustness of the hyperentangled state against depolarizing noise. The witness operator used here is the same as the witness of bit flip channel Eq.(\ref{eq:witness}). As the witness is the same, we are able to use the same decomposition in Eq.(\ref{eq:witness_decomp}) for it. In Fig.\,\ref{fig:witnessdepol}, $\langle W \rangle$ for the $\rho^{out}_{E}$ can be seen to approach $\langle W\rangle = 0$ with increasing noise levels and eventually crosses it at the same level of noise for which $\mathcal{N} = 0$, indicating that the state is now separable. But for $\rho^{out}_{HE}$, $\langle W\rangle<0$ throughout the noise levels. This agrees with the corresponding entanglement negativity plot of the depolarizing channel.

\begin{figure}[h!]
  \raggedleft
\includegraphics[scale = 0.32]{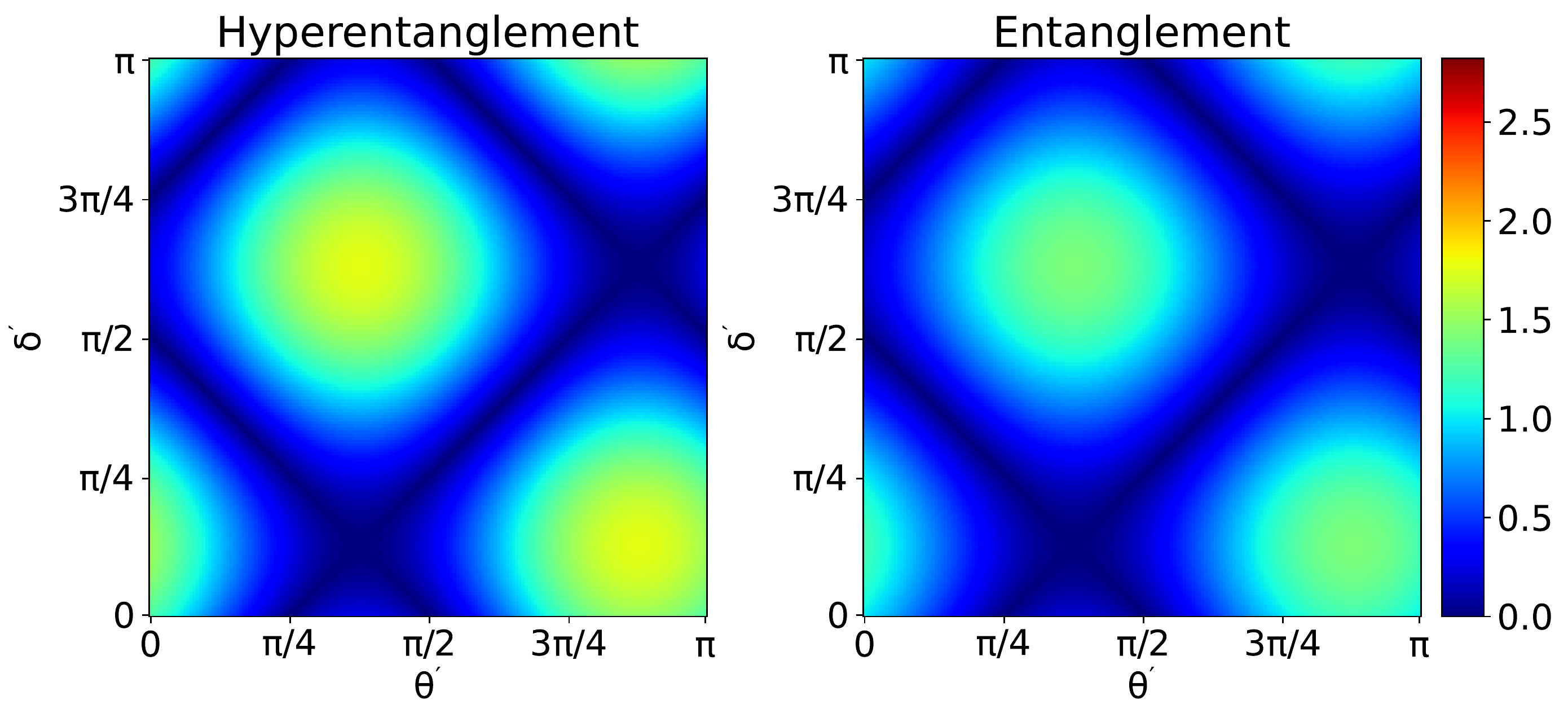}
\caption{CHSH parameter $S$ as a function of $\theta^{\prime}$ and $\delta^{\prime}$ when $(\theta, \delta)  =(\pi/4, \pi/2)$ and the depolarizing noise level is set to $p = 0.5$. Both the plots have a similar pattern with a slightly lower $S$ value for the entangled state.}
 \label{fig:depolnon} 
\end{figure}

\begin{figure}[h!]
\includegraphics{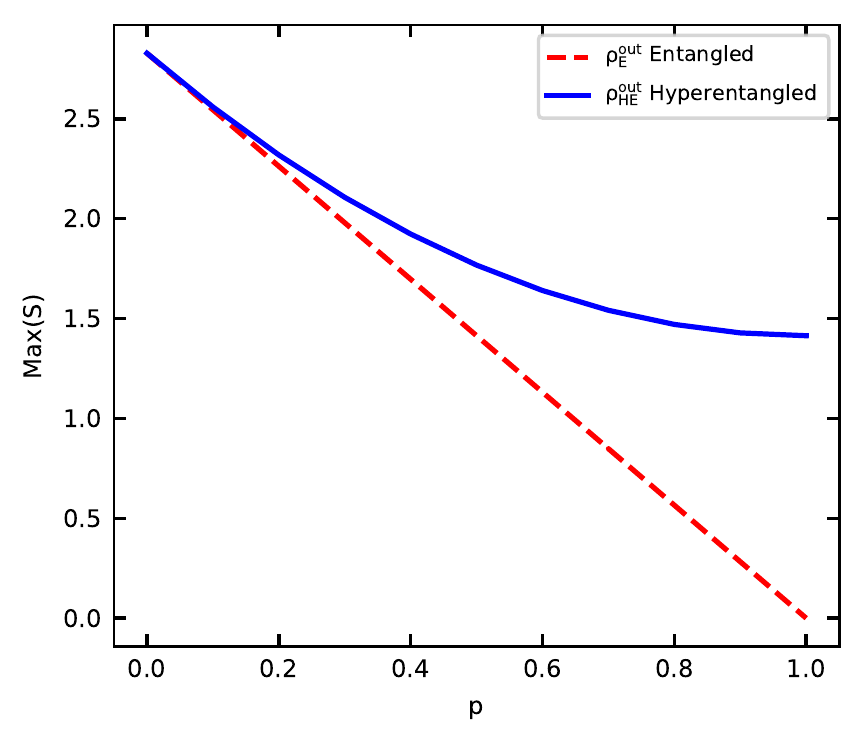}
\caption{Maximum value of CHSH parameter $S$ as a function of depolarizing noise level $p$ when $(\theta, \delta) = (\pi/2, \pi/4)$ and $\theta',\delta' \in [0,\pi]$. The plot for the hyperentangled state shows a higher level of Max($S$) compared to the entangled state. } \label{fig:depolnonplot} 
\end{figure}

The nonlocality plots, $S$ and Max$(S)$ in Fig.\,\ref{fig:depolnon} and Fig.\,\ref{fig:depolnonplot}, respectively also show a clear advantage for hyperentanglement, indicating enhanced retention of nonlocality through depolarizing channel.
%================================
\subsection{Phase Damping Noise}
%================================

\begin{figure}[h!]
 \includegraphics{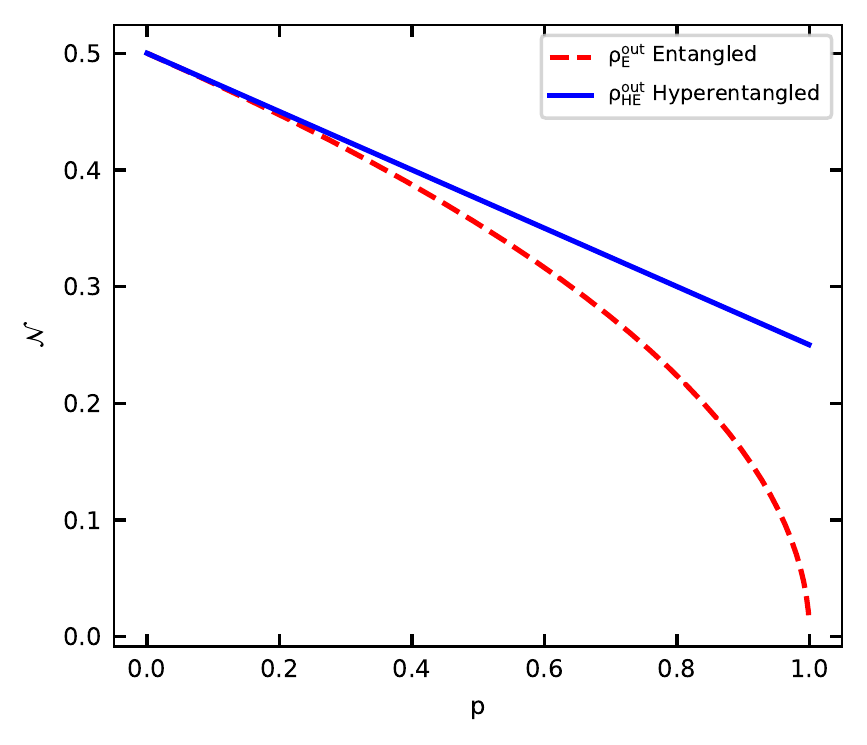} 
\caption{Entanglement negativity $\mathcal{N}$ as a function of phase damping noise level $p$. The negativity for the hyperentangled state is slightly higher compared to the entangled state with an increase in the difference for higher levels of noise.}
\label{fig:phasecontrol}
\end{figure}

\begin{figure}[h!]
  \includegraphics{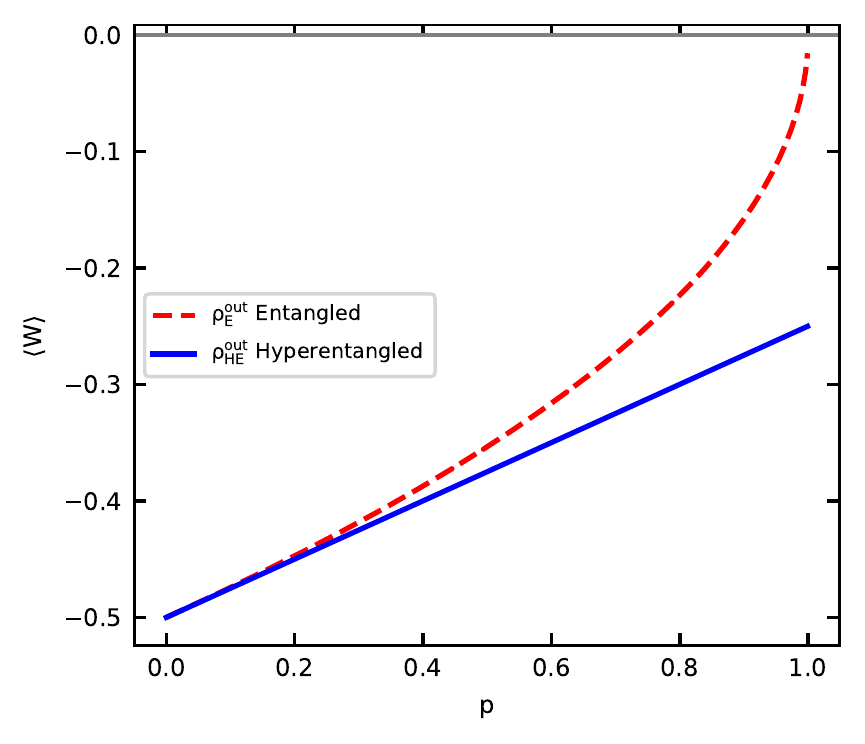}
 \caption{Entanglement witness $\langle W \rangle$ as a function of phase damping noise level $p$. The plots are consistent with the negativity plot Fig.\,\ref{fig:phasecontrol} showing a similar inverse trend as seen for other noise channels.}
 \label{fig:witnessphase}
\end{figure}

The phase damping noise channel is used to model the loss of a fixed relative phase between the states of a quantum system, due to interactions with the environment. This process also termed as decoherence, is a frequently encountered effect in photons. The Kraus operators for the phase damping noise model are,
\begin{equation}
K_1 = \begin{pmatrix}
1 & 0 \\
0 & \sqrt{1-p)}
\end{pmatrix}
\quad
K_2 = \begin{pmatrix}
0 & 0 \\
0 & \sqrt{p} 
\end{pmatrix}.
\end{equation}
In Fig.\,\ref{fig:phasecontrol} and Fig.\,\ref{fig:witnessphase} the negativity and witness as a function of phase damping noise level is shown. Hyperentangled states shows a slight advantage against phase damping noise as compared to the entangled state.

\begin{figure}[h!]
 \centering
\includegraphics[scale = 0.32]{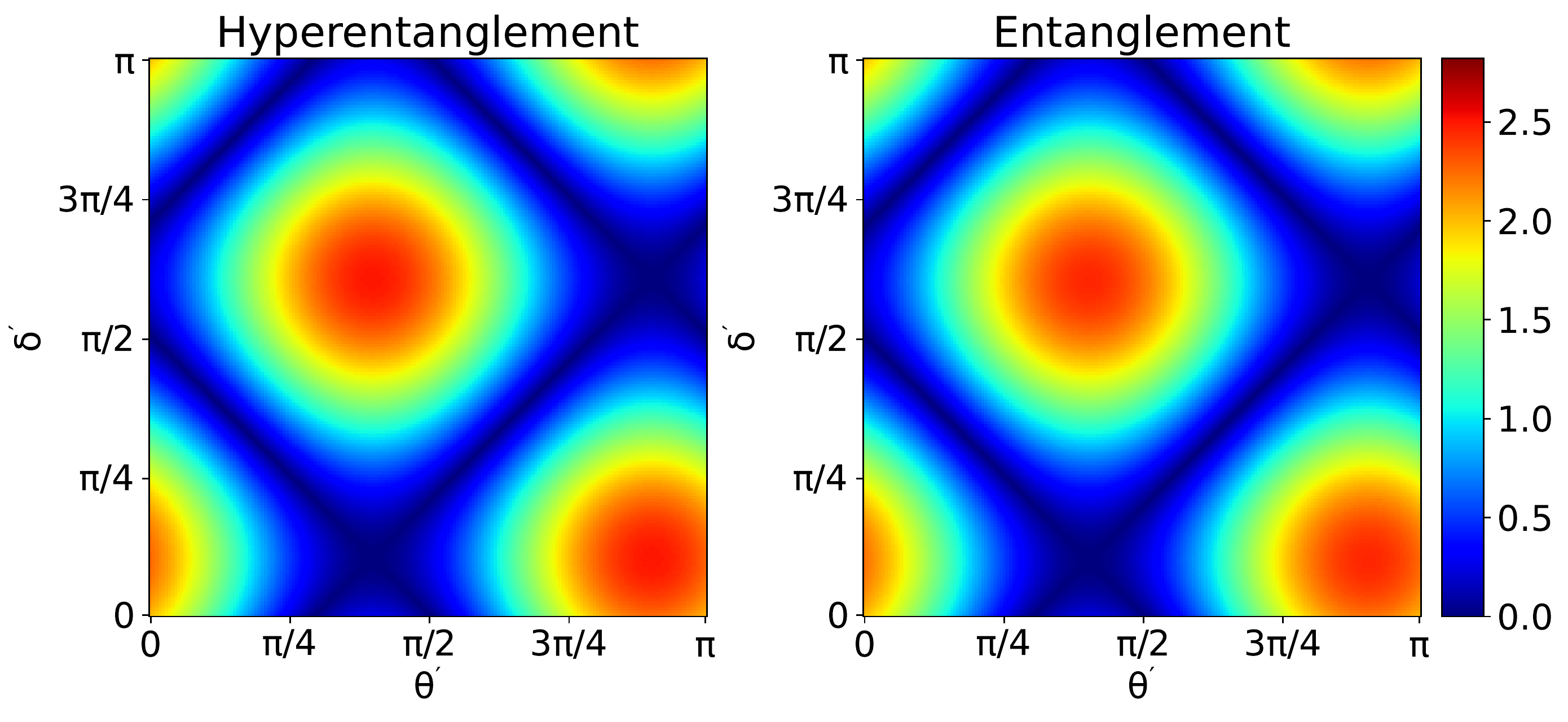}
\caption{Nonlocality plots for the phase damping channel with p = 0.5. $\theta = \pi/4$ and $\delta = \pi/2$ is fixed and $\theta'$ and $\delta'$ are varied. Both plots show a negligible difference compared to each other and the zero noise case Fig.\,\ref{fig:refnon}.}
 \label{fig:phasenon} 
\end{figure}

However, for the phase damping noise channels, it can be seen that both the states undergoing phase damping noise still violate the CHSH inequality. The plots obtained in Fig.\,\ref{fig:phasenon} indicates that the phase damping channel has a negligible effect on the nonlocality of a quantum state.

As seen in the nonlocality plots for the phase damping noise channel, the S parameter remains above 2 for all the noise levels except at p = 1 for the entangled state. For the entangled state, the phase damping channel with p = 1 decoheres it to a classical state which saturates the CHSH inequality, S = 2.
\begin{figure}[h!]
\includegraphics{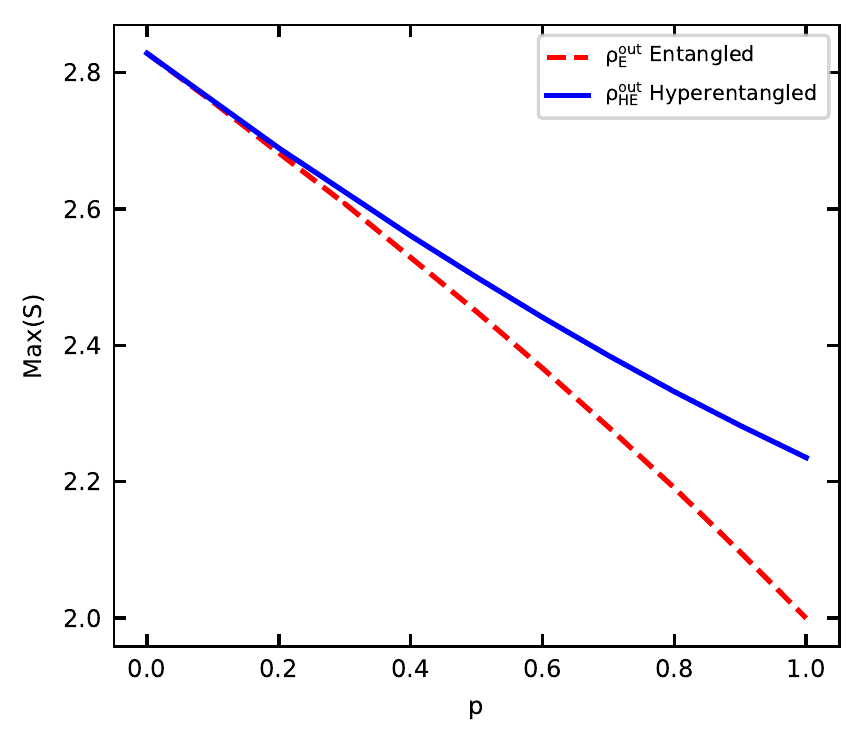}
\caption{Maximum value of CHSH parameter as a function of Phase damping noise level p when $\theta = \pi/2$ and $\delta = \pi/4$ and $\theta',\delta' \in [0,\pi]$. Both the plots show a downtrend, but the maximum S values still violate CHSH inequality for both the plots.}
 \label{fig:phasenonplot} 
\end{figure} 
In the case of the phase damping noise, there appears to be only a minor difference between the nonlocality of $\rho^{out}_{(E)}$ and $\rho^{out}_{(HE)}$. This difference may possibly be negligible to be detected in a laboratory setting. 
\subsection{Unitary Noise Model}
The Kraus operators of a noise channel can be used to back calculate an equivalent unitary that acts on a system coupled with the environment. This unitary gives equivalent dynamics to the noise channel when the environment is traced out.
This method termed as stinespring dilation is given by \cite{Stinespring1955Positive},
\begin{equation}
  U_{A\rightarrow A'E}: \ket{\psi} \rightarrow \sum_{a} K_{a} \ket{\psi} \otimes \ket{a},
  \label{eq:stine} 
\end{equation}
where E corresponds to the environment coupled with the system A with dimension equal to the number of Kraus operators and orthonormal basis $ \{\ket{a}\} $. 

We construct a dilated unitary for the bit flip channel, 
\begin{equation}
  \label{eq:dilated} 
  U_x = \begin{pmatrix}
    \sqrt{1-p} & 0 & 0 & \sqrt{p} \\
    0 & \sqrt{1-p} & \sqrt{p} & 0 \\
    0 & -\sqrt{p} & \sqrt{1-p} & 0 \\
    -\sqrt{p} & 0 & 0 & \sqrt{1-p}
  \end{pmatrix}.
\end{equation}

Now we model the noise using a controlled unitary given by,
\begin{align}
  U =& P_{00}\otimes (U_x \otimes U_x) + P_{01} \otimes (U_x \otimes I),\\ 
     &+ P_{10} \otimes (I \otimes U_x) + P_{11} \otimes (I \otimes I).\nonumber
\label{eq:CUni} 
\end{align}
Here U acts on a system in which the polarization subsystems of the photons are both coupled with environment systems. For example, the polarization entangled state is coupled with the environment as follows, 

\begin{align}
(1/\sqrt{2})(\ket{HV}+ \ket{VH}) \longrightarrow  (1/\sqrt{2})(&\ket{H}\ket{0}\ket{V}\ket{0} \\
+ &\ket{V}\ket{0}\ket{H}\ket{0}),\nonumber 
\end{align}

where $\ket{0}$  is the initial state of the environment. After the evolution of the environment-coupled hyperentangled and entangled states with U, the environment can now be traced out to obtain the final state. The correlations in the path and polarization subsystems can be studied by tracing each other out respectively. Here we plot the negativity of each subsystems as a function of the noise level $p$.

\begin{figure}[h!]
\includegraphics{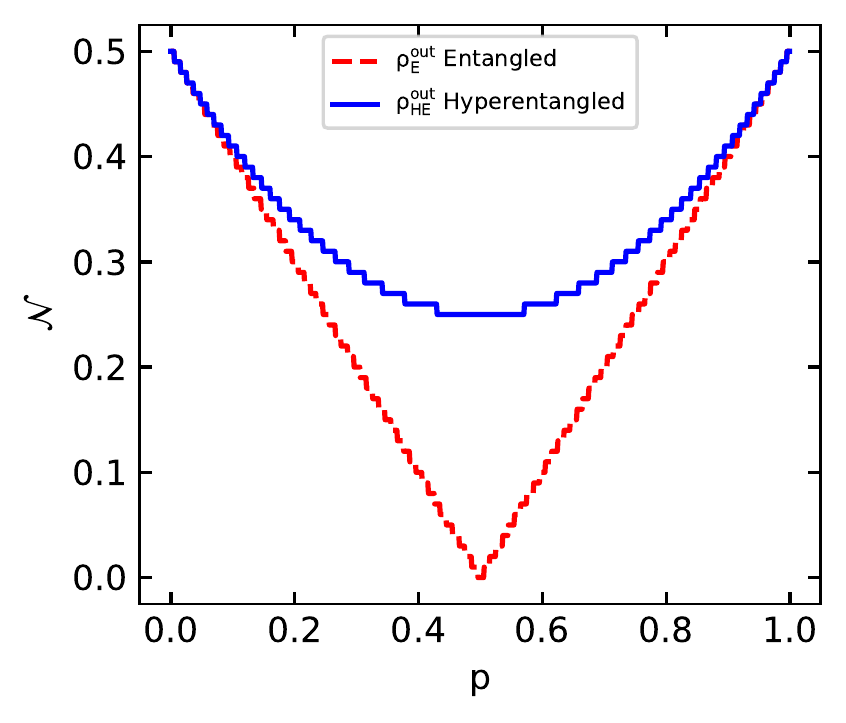}
\caption{Negativity in the polarization degree of freedom as a function of bit flip noise level p using Stinespring dilation. The plots are equivalent to the negativity plots obtained for the bit flip channel Fig.\,\ref{fig:blipcontrol} showing the robustness of hyperentanglement.} 
 \label{fig:blipplotI} 
\end{figure} 
\begin{figure}[h!]
\includegraphics{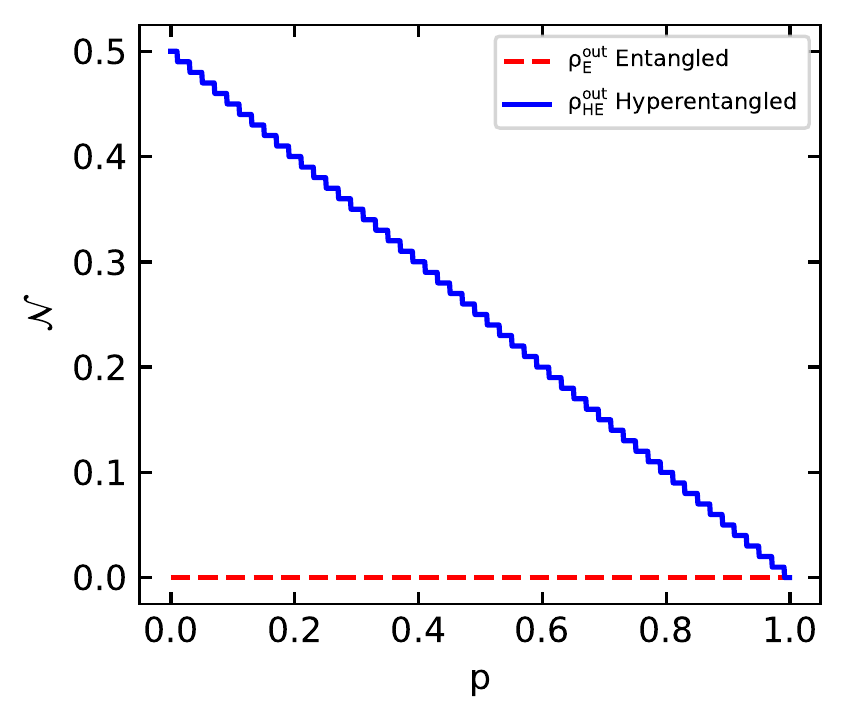}
\caption{Negativity in the path degree of freedom as a function of bit flip noise level p using Stinespring dilation. The negativity in the entangled state remains at 0 owing to no path entanglement, and the negativity for the hyperentangled state shows an inverse linear trend with the noise level.}
 \label{fig:blipplotII} 
\end{figure} 
The plot of negativity of polarization degree of freedom in Fig.\,\ref{fig:blipplotI} is identical to the plot for negativity in Fig.\,\ref{fig:blipcontrol} except for the small fluctuations in the curves which can be attributed to the small crossover of correlation of one degree of freedom to the other. In Fig\,\ref{fig:blipplotII}, we observe an inverse linear relation of the negativity in the path subsytem of the hyperentangled state and the noise level $p$. The negativity in the path of the conventional entangled state remains at zero as there is no path entanglement present in it. This is an additional feature we are able to probe because of the unitary noise model compared to using Kraus operators. From Fig.\,\ref{fig:blipplotII} it is evident that some correlations in the paths do make across through the noise.

%==================================
\section{CONCLUSION}
\label{conc}
%==================================

The noise model proposed in this paper serves as a useful tool to theoretically study the effect of various types of noises encountered in the experimental settings where one of the basis state of an entangled pair is subject to noise.  One of the main features of quantum systems that sets them apart from classical systems are the quantum correlations which have been proven to be a useful resource in many applications. Any approach to protect the entanglement in quantum systems from environmental effects will always be very useful.  Higher dimensional entanglement in quantum systems has proven its advantages over conventional entanglement in many applications like dense coding and quantum illumination. The results obtained here for the path-polarization hyperentangled state of photons pairs help confirm this advantage and possibly provides a reason for it. The entanglement in the path, provides a way to retain better correlations through a noisy environment. The hyperentangled states are shown to have enhanced negativity as compared to the conventional entangled state of photons. An experimental method to confirm this is provided in the form of entanglement witnesses, that provide a method to verify whether a state is entangled or not using a few joint local measurements. The CHSH parameter violation is also computed and it shows that hyperentangled states retain nonlocality as good as, or better than entangled states. It can be concluded that path-polarization hyperentangled state Eq.~(\ref{eq:Hyperentangled}) can be used as a robust probe in a noisy environment for applications like quantum communication and quantum illumination.  Though we have shown the robustness for a specific path-polarization hyper entangled state,  the conclusion will hold in general for all forms of path-polarization hyper entangled states and should be extendable for all hyperentangled states. 

%====================================================================

 \vskip 0.2in
\noindent \textit{\bf Acknowledgement.}We acknowledge the support from the Office of Principal Scientific Advisor to Government of India, project no. Prn.SA/QSim/2020.
%===========================================================


\begin{thebibliography}{4}
%Intro linea
\bibitem{Entanglement_Horodecki}
R. Horodecki, P. Horodecki, M. Horodecki, and K. Horodecki, Quantum entanglement, \href{https://doi.org/10.1103/RevModPhys.81.865}{Rev. Mod. Phys. \textbf{81}, 865 (2009)}.

\bibitem{Advances2011Giovanetti}
V. Giovannetti, S. Lloyd, and L. Maccone, Advances in quantum metrology, \href{https://doi.org/10.1038/nphoton.2011.35}{Nat. Photonics \textbf{5}, 222–229 (2011) }.
%Entanglement enhancements
\bibitem{E91} A. Ekert, Quantum cryptography based on Bell’s theorem, \href{https://doi.org/10.1103/PhysRevLett.67.661}{Phys. Rev. Lett. \textbf{67}, 661 (1991)}.

\bibitem{Bennet1992Dense} C. H. Bennett and S. J. Wiesner, Communication via one- and two-particle operators on Einstein-Podolsky-Rosen states,\href{https://doi.org/10.1103/PhysRevLett.69.2881}{Phys. Rev. Lett. \textbf{69}, 2881 (1992)}.

\bibitem{LLoyd_QI2008}
 S. Lloyd, Enhanced Sensitivity of Photodetection via Quantum Illumination, \href{https://doi.org/10.1126/science.1160627}{Science \textbf{321} 5895, (2008)}.

\bibitem{Teleporting1993Bennett} C. H. Bennett, G. Brassard, C. Crépeau, R. Jozsa, A. Peres, and W. K. Wootters, Teleporting an unknown quantum state via dual classical and Einstein-Podolsky-Rosen channels, \href{https://doi.org/10.1103/PhysRevLett.70.1895}{Phys. Rev. Lett. \textbf{70}, 1895 (1993)}.

\bibitem{Josza2003Role} R. Jozsa and N. Linden, On the role of entanglement in quantum-computational speed-up, \href{https://doi.org/10.1098/rspa.2002.1097}{Proc. Math. Phys. Eng. Sci. \textbf{459}, 2036 (2003)}.

\bibitem{Acin2007Deviceindependent} A. Acín, N. Brunner, N. Gisin, S. Massar, S. Pironio, and V. Scarani, Device-Independent Security of Quantum Cryptography against Collective Attacks, \href{https://doi.org/10.1103/PhysRevLett.98.230501}{Phys. Rev. Lett. \textbf{98}, 230501 (2007)}.
%%%%

\bibitem{Hu2020Beating}
X.-M. Hu, Y. Guo, B.-H. Liu, Y.-F. Huang, C.-F. Li, and G.-C. Guo, Beating the channel capacity limit for superdense coding with entangled ququarts,\href{https://doi.org/10.1126/sciadv.aat9304}{Sci. Adv. \textbf{6}, 37, (2020)}.


\bibitem{Cerf2002Security}
N. J. Cerf, M. Bourennane, A. Karlsson, and N. Gisin, Security of Quantum Key Distribution Using d-Level Systems, \href{https://doi.org/10.1103/PhysRevLett.88.127902}{Phys. Rev. Lett. \textbf{88}, 127902 (2002)}.

\bibitem{Zhu2020IsHigh}
F. Zhu, M. Tyler, N. H. Valencia, M. Malik, and J. Leach, Is high-dimensional photonic entanglement robust to noise? \href{https://doi.org/10.1116/5.0033889}{AVS Quantum Sci. \textbf{3}, 011401 (2021)}.

\bibitem{Collins2002Arbitrary}
D. Collins, N. Gisin, N. Linden, S. Massar, and S. Popescu, Bell Inequalities for Arbitrarily High-Dimensional Systems, \href{https://doi.org/10.1103/PhysRevLett.88.040404}{Phys. Rev. Lett. \textbf{88}, 040404 (2002)}.

\bibitem{NewSourceEntanglement_Kwiat1995} P.G Kwiat., K. Mattle, H. Weinfurter, A. Zeilinger, A.V. Sergienko, and Y. Shih, New High-Intensity Source of Polarization-Entangled Photon Pairs, \href{https://doi.org/10.1103/PhysRevLett.75.4337}{Phys. Rev. Lett. \textbf{75}, 4337 (1995)}.


%%%%
\bibitem{Chen2017Exact}
Z. Chen, Y. Zhou, and J.-T. Shen, Exact dissipation model for arbitrary photonic Fock state transport in waveguide QED systems, \href{https://doi.org/10.1364/OL.42.000887}{Opt. Lett. \textbf{42} (4), 887 (2017)}. 

\bibitem{Chen2018Entanglement}
Z. Chen, Y. Zhou, and J.-T, Entanglement-preserving approach for reservoir-induced photonic dissipation in waveguide QED systems, \href{https://doi.org/10.1103/PhysRevA.98.053830}{Phys. Rev. A \textbf{98}, 053830 (2018)}.

\bibitem{Chen2017Dissipation}
Z. Chen, Y. Zhou, and J.-T, Dissipation-induced photonic-correlation transition in waveguide-QED systems, \href{https://doi.org/10.1103/PhysRevA.96.053805}{Phys. Rev. A \textbf{96}, 053805 (2017)}.

%Generation of Ent, Hyperent
 
\bibitem{Kwiat1996Hyperentanglement}
P. G. Kwiat, Hyper-entangled states, \href{https://doi.org/10.1080/09500349708231877}{J. Mod. Opt. \textbf{44} 2173 (1997)}. 

\bibitem{BCM05} M. Barbieri, C. Cinelli, P. Mataloni, and F. De Martini, Polarization-momentum hyperentangled states: Realization and characterization, \href{https://doi.org/10.1103/PhysRevA.72.052110}{Phys. Rev. A {\bf 72} 052110 (2005)}.

\bibitem{BMM06} M. Barbieri, F. D. Martini, P. Mataloni, G. Vallone, and A. Cabello,  Enhancing the Violation of the Einstein-Podolsky-Rosen Local Realism by Quantum Hyperentanglement, \href{https://doi.org/10.1103/PhysRevLett.97.140407}{Phys. Rev. Lett. {\bf 97}, 140407 (2006).}

\bibitem{GenerationHyper_Barreiro2005}
J. T. Barreiro, N. K. Langford, N. A. Peters, and P. G. Kwiat, Generation of Hyperentangled Photon Pairs, \href{https://doi.org/10.1103/PhysRevLett.95.260501}{Phys. Rev. Lett. \textbf{95}, 260501 (2005)}.


\bibitem{Zhao2019Direct}
T.-M. Zhao, Y. S. Ihn, and Y.-H. Kim, Direct Generation of Narrow-band Hyperentangled Photons, \href{https://doi.org/10.1103/PhysRevLett.122.123607}{Phys. Rev. Lett. \textbf{122}, 123607 (2019)}.

\bibitem{Dong2014Generation}
S. Dong et al., Generation of hyper-entanglement in polarization/energy-time and discrete-frequency/energy-time in optical fibers, \href{https://doi.org/10.1038/srep09195}{Sci. Rep. \textbf{5}, 9195 (2015)}.

\bibitem{Yaasir2021Generation}
P. A. A. Yasir and C. M. Chandrashekar, Generation of hyperentangled states and two-dimensional quantum walks using J or q plates and polarization beam splitters, \href{https://doi.org/10.1103/PhysRevA.105.012417}{Phys. Rev. A \textbf{105}, 012417 (2022)}.

\bibitem{PMHyper_Barbieri2005}
M. Barbieri, C. Cinelli, P. Mataloni, and F. De Martini, Polarization-momentum hyperentangled states: Realization and characterization, \href{https://doi.org/10.1103/PhysRevA.72.052110}{Phys. Rev. A \textbf{72}, 052110 (2005)}.

%Hyperentanglement enhancements

\bibitem{Barreiro2008Beating}
J.T. Barreiro, T.C. Wei, and P.G. Kwiat, Beating the channel capacity limit for linear photonic superdense coding, \href{https://doi.org/10.1038/nphys919}{Nature Physics \textbf{4}, pages 282–286 (2008)}.

\bibitem{Sheng2021Onestep}
Y.-B. Sheng, L. Zhou, and G.-L. Long, One-step quantum secure direct communication, \href{https://doi.org/10.1016/j.scib.2021.11.002}{Sci. Bull. \textbf{67} pp 367-374 (2021)}.

\bibitem{Hu2021LongDistance}
X.-M. Hu et al., Long-Distance Entanglement Purification for Quantum Communication, \href{https://doi.org/10.1103/PhysRevLett.126.010503}{Phys. Rev. Lett. \textbf{126}, 010503 (2021)}.

\bibitem{Huang2022Experimental}
C.-X. Huang et al., Experimental one-step deterministic polarization entanglement purification, \href{https://doi.org/10.1016/j.scib.2021.12.018}{Sci. Bull. \textbf{67} 593-597 (2022)}.

\bibitem{HyperentanglementQI_AVPrabhu2021} A.V. Prabhu, B.  Suri, and C.M Chandrashekar, Hyperentanglement-enhanced quantum illumination, \href{https://doi.org/10.1103/PhysRevA.103.052608}{Phys. Rev. A \textbf{103}, 052608 (2021)}.

\bibitem{nielsenchuang}
M. A. Nielsen and I. L. Chuang, Quantum computation and quantum information. Cambridge Cambridge University Press, 2019.

\bibitem{mmwilde_qiqc}
M. Wilde, Quantum information theory.Cambridge University Press, 2013.

\bibitem{Simon2002Polarization}
C. Simon and J.-W. Pan, Polarization Entanglement Purification using Spatial Entanglement, \href{https://doi.org/10.1103/PhysRevLett.89.257901}{Phys. Rev. Lett. 89, 257901 (2002)}.

\bibitem{Olivier2001Discord}
H. Ollivier and W. H. Zurek, Quantum Discord: A Measure of the Quantumness of Correlations,\href{https://doi.org/10.1103/PhysRevLett.88.017901}{Phys. Rev. Lett. \textbf{88}, 017901 (2001)}.

\bibitem{Hill1997Entanglement}
S. Hill and W. K. Wootters, Entanglement of a Pair of Quantum Bits, \href{https://doi.org/10.1103/PhysRevLett.78.5022}{Phys. Rev. Lett. \textbf{78}, 5022 (1997)}.


\bibitem{ComputableEntanglment_Vidal2002}
G. Vidal and R. F. Werner, Computable measure of entanglement, \href{https://doi.org/10.1103/PhysRevA.65.032314}{Phys. Rev. A \textbf{65}, 032314 (2002)}.

\bibitem{VolumeofSeprablestates_Zyckowski1999}
K. Życzkowski, P. Horodecki, A. Sanpera, and M. Lewenstein, Volume of the set of separable states, \href{https://doi.org/10.1103/PhysRevA.58.883}{Phys. Rev. A \textbf{58}, 883 (1999)}.

\bibitem{QST_DanielFV2001}
D. F. V. James, P. G. Kwiat, W. J. Munro, and A. G. White, Measurement of qubits, \href{https://doi.org/10.1103/PhysRevA.64.052312}{Phys. Rev. A \textbf{64}, 052312 (2001)}.

\bibitem{EntanglementWitness_Terhal2000}
B. M. Terhal, Bell inequalities and the separability criterion, \href{https://doi.org/10.1016/S0375-9601(00)00401-1}{Physical Letters A \textbf{271} 5-6 (2000)}.

\bibitem{DetectionofEntanglment_Guhne2002} 
O. Guehne, P. Hyllus, D. Bruss, A. Ekert, M. Lewenstein, C. Macchiavello, A. Sanpera, Detection of entanglement with few local measurements, \href{https://doi.org/10.1103/PhysRevA.66.062305}{Phys. Rev. A \textbf{66}, 062305 (2002)}.

\bibitem{Bell1964}
J. S. Bell, On the Einstein Podolsky Rosen paradox, \href{https://doi.org/10.1103/PhysicsPhysiqueFizika.1.195}{Physics Physique Fizika \textbf{1}, 195 (1964)}.

\bibitem{CHSH_aspect1981}
A. Aspect, P. Grangier, and G. Roger, Experimental Tests of Realistic Local Theories via Bell’s Theorem, \href{https://doi.org/10.1103/PhysRevLett.47.460}{Phys. Rev. Lett. \textbf{47}, 460 (1981)}.

\bibitem{CHSH_1970}
J. F. Clauser, M. A. Horne, A. Shimony, and R. A. Holt, Proposed Experiment to Test Local Hidden Variable Theories., \href{https://doi.org/10.1103/PhysRevLett.23.880}{Phys. Rev. Lett. \textbf{23}, 880 (1970)}.

\bibitem{Stinespring1955Positive}
W. F. Stinespring, Positive functions on C*-algebras, \href{https://doi.org/10.1090/S0002-9939-1955-0069403-4}{Proc. Amer. Math. Soc. \textbf{6}, 211-216 (1955)}.
\end{thebibliography}
\end{document}